\documentclass[12pt]{article}
\usepackage{amsmath}
\usepackage{graphicx,psfrag,epsf}
\usepackage{enumerate}
\usepackage{natbib}
\usepackage{url}
\usepackage{hyperref}
\usepackage{float}
\usepackage{bm}
\usepackage{multirow}
\usepackage{setspace}
\usepackage{appendix}

\newcommand{\blind}{0}  

\usepackage{natbib}

\newcommand{\omb}{{\boldsymbol{\omega}}}
\newcommand{\alb}{{\boldsymbol{\alpha}}}
\newcommand{\ombp}{{\boldsymbol{\psi}}}
\newcommand{\thetab}{{\boldsymbol{\theta}}}
\newcommand{\xib}{{\boldsymbol{\xi}}}

\newcommand{\Ndark}{N_{\emptyset}^{}}
\newcommand{\Thetabm}{\Theta \setminus \thetab}
\newcommand{\mut}{\hat{\mu}_\thetab^*[\Thetabm]}
\newcommand{\xbm}{\mathbf{x}}
\newcommand{\tbm}{\mathbf{t}}
\newcommand{\Abm}{\mathbf{A}}
\newcommand{\SparseMSE}{{\tt SparseMSE}}
\newcommand{\SparseMSEcitep}{ \citep{SparseMSE}}

\newtheorem{proposition}{Proposition}

\begin{document}

%


\title{\bf Multiple Systems Estimation for Sparse Capture Data: Inferential Challenges when there are Non-Overlapping Lists}
\if0\blind
{
\author{Lax Chan, Bernard W. Silverman  and Kyle Vincent    \\   Rights Lab, University of Nottingham, U.K.
\thanks{This work was supported by the Arts and Humanities Research Council and the Economic and Social Research Council [grant number ES/P001491/1] grant Modern Slavery: Meaning and Measurement (PaCCS Transnational Organised Crime, University of Nottingham, 2016--18).}}
} \fi
\maketitle
\begin{abstract}
Multiple systems estimation strategies have recently been applied to quantify hard-to-reach populations, particularly when estimating the number of victims of human trafficking and modern slavery. In such contexts, it is not uncommon to see sparse or even no overlap between some of the lists on which the estimates are based.
These create difficulties in model fitting and selection, and we develop inference procedures to address these challenges. The approach is based on Poisson log-linear regression modeling. Issues investigated in detail include taking proper account of data sparsity in the estimation procedure, as well as the existence and identifiability of maximum likelihood estimates.  A stepwise method for choosing the most suitable parameters is developed, together with a bootstrap approach to finding confidence intervals for the total population size. We apply the strategy to two empirical data sets of trafficking in US regions, and find that the approach results in stable, reasonable estimates. An accompanying R software implementation has been made publicly available. Supplementary materials for this paper are available online.
\end{abstract}

\noindent%
{\it Keywords:}   Human trafficking; Log-linear models; Mark-recapture; Model identifiability; Model selection;  Modern slavery; Poisson regression modeling.
\newpage
\section{Introduction}

Multiple systems estimation, a generalization of the mark-recapture approach \citep{Petersen1896, Schwarz1999}
is a class of methods that can be used to estimate the size of hard-to-reach populations in many contexts, including, in recent years, those comprised of human trafficking or slavery victims.      The methods are typically applied to wildlife populations \citep{Williams2002} and to hidden populations such as injection drug users \citep{King2013}.  In the administrative or law enforcement context, multiple systems estimation aims to read across from lists of observed or identified individuals from a study population to estimate the total population of interest; see, for example, \cite{Bales2015} and \cite{Cruyff2017}.  A mathematical model is posited for the pattern of incidences across the lists, and the ``dark figure", the number of unobserved cases, is estimated. A survey of the history of the methods and a range of applications is provided, for example, by \cite{Bird2018}.

Because the method estimates the number of victims including those that are not directly observed or detected, it plays an especially important role in policy making decisions to help combat human trafficking and modern slavery.  For example, as set out in \cite{Bales2015}, a multiple systems estimate constructed from data collated by a government agency was a key component of the strategy \citep{HomeOffice2014} leading to the UK Modern Slavery Act 2015.

A frequent specific challenge posed by data on human trafficking is sparse overlap between the observed administrative lists; indeed, it appears to be the norm rather than the exception that there will be pairs of lists between which there is no observed overlap.  This sparsity can lead to inferential and algorithmic difficulties and instabilities if it is not addressed.  In applications such as wildlife populations, the researcher may be able to continue capturing from the study population until sufficient overlap is observed between the capture occasions.  Such a strategy is not available in the human trafficking context, nor usually in other human rights areas either.  

A pair of lists may fail to overlap for a number of reasons:  there may be a genuine structural reason why the particular lists cannot overlap; there may be negative correlation between lists; or it may simply be that the overall sample size is relatively small and, especially if the two lists have small capture probabilities, there do not happen to be any cases which are on both lists.  In this area, there is as yet limited understanding of data and of mechanisms, and furthermore data are often highly anonymized for reasons of confidentiality and security.  Typically, those analyzing the data may not know anything about a list other than an uninformative label, because the collation between lists is carried out by a single trusted individual or agency on that understanding.   See for example \cite{Bales2015} and \cite{Bales2018}.
Hence, there may be no further information available, beyond simply the number of overlapping cases, as to why no cases are observed in common between two lists.

%

We approach inference via Poisson log-linear regression modeling applied to counts of individuals that are observed on each possible combination of the lists. 
This is a well-known technique that allows one to model correlations and dependencies between lists.  
The standard approach is set out by \cite{San1984}, \cite{Cormack1989}, \cite{Cormack1992}, \cite{Rivest2004} and \cite{Bird2018}, among others, and implemented in \cite{Baillargeon2007} and \cite{Rivest2012}. However, as \cite{Fienberg2012} discuss in a much more general context, contingency tables with zero entries, as will arise if there is sparse overlap between lists, may lead to cases where to carry out maximum likelihood estimation it is necessary to extend the range of parameters to include $- \infty$, and even then the maximum likelihood estimate of the model parameters may not exist or may not be identifiable.

In our context, therefore, empty overlaps between lists require careful treatment.  The primary objective of this paper is to introduce inferential procedures and computational implementations that explicitly handle this case. For simplicity, we have focused attention only on models which include parameters for two-list effects (also called `first-order interactions' by some authors), but the basic concepts of allowing for empty cells, and of checking for the existence of estimates, are straightforward to extend.

We first of all develop a method which fits a model stably, taking proper account of existence and identifiability issues that can arise if the data are sparse.  We then consider a model-selection procedure to choose the most suitable set of parameters on which to base inference.  A stepwise approach to model selection is used, but so that any effects of choosing the specific model are taken into account in the inference, confidence intervals for the estimation are constructed using the $BC_a$ bootstrap method \citep{DiCEfr96}.

The methods are motivated and illustrated by data sets based on human trafficking victims in the New Orleans area \citep{Bales2018} and the Western site of a research study in the USA \citep{Farrell2019}.    Simulations studies are used to validate the stepwise approach and are based on data sets generated from these empirical data sets, where multinomial sampling is use to assign capture histories to the population members as suggested by \cite{Cormack1992}.  
We conduct our analyses in the R programming language \citep{Rprogram}, and have developed an accompanying R software package \SparseMSE \SparseMSEcitep. The package allows readers to implement the methodology on their own data as well as to reproduce the results presented in the paper.

The paper is organized as follows. Section 2 outlines the Poisson log-linear model and gives the notation and likelihood setup, details specific issues concerning the existence and identifiability of maximum likelihood estimates, and also discusses issues relating to the breakdown of the assumptions underlying standard likelihood-ratio and information-theoretic approaches.  It also develops algorithms for checking efficiently whether models present problems of nonexistence or unidentifiability of estimates.  Section 3 develops the model-selection routine and corresponding inference procedure, setting out an efficient algorithmic approach to the bootstrap in this case.  A simulation comparison with one of the current standard methods is included.  Section 4 presents the results from the two empirical applications, as well as a simulation study informing the choice of threshold in our procedure.  The concluding remarks in Section 5 include a comment on the R package, as well as discussions of the possible extension of the procedure to higher-order interactions, and to data with covariate information.  There is further detail of several topics in the supplementary materials to the paper.  

\section{The Model}
\label{sec:modeldef}

We first define notation and set out the model. The framework leads to an algorithmic approach facilitating correct and stable calculations. We then discuss the implications of sparse counts on existing inferential methods, followed by a discussion on checks for existence of maximum likelihood estimates and identifiability of the model.

\subsection{Notation and Definitions}
Suppose we have $t$ capture occasions, or lists, on which members of the population can occur.
An individual's {\em capture history} is the set of lists on which the individual is actually observed, or captured.
A capture history is a subset $\omb$ of $\{1, 2, \ldots, t\}$.

Now suppose that there are $m$ individuals captured at least once in our study.  Denote the $m$ observed capture histories by $\xib_1, \xib_2, \ldots, \xib_m$.
For any particular capture history $\omb$ define $N_\omb^{}$ to be the number of individuals observed to have exactly that capture history, i.e. the number of $\xib_i$ equal to $\omb$.
It is important to note that the actual data consist of a sample of size $m$ from a discrete distribution over the possible capture histories.

The {\em order} of a capture history is defined to be the number of captures in the set.
The braces are often omitted when the members of the history are given as suffices.   Thus, for example, if $t = 4$ the capture history $\{ 1, 3 \}$ has order 2, and $N_{\{ 1, 3 \}}^{}$, usually written $N_{13}^{}$, 
is the number of individuals which are observed on both lists 1 and 3 but not on lists 2 or 4.
A particular capture history of interest, with order 0, is the {\em null capture history} $\emptyset$.
The quantity $\Ndark$ is the {\em dark figure} of individuals which are not captured on any list, and therefore cannot be observed.
The observed data give rise to the $2^t-1$ values $\{N_\omb^{} : \omb \ne \emptyset\}$, which we will also write as $\bf N$.


It is characteristic of data collected in the modern slavery context that there will be some capture histories for which the observed count is zero.  Typically, each list only records a relatively small proportion of the total population, because of the ``hidden'' nature of modern slavery as a crime, and the numbers of cases recorded on any particular pattern of overlaps between lists can easily be considerably smaller.



For any capture history $\omb$ define
\begin{equation}
\label{eq:nstardef}
N^*_\omb = \sum_{\ombp \supseteq \omb, \ombp \ne \emptyset} N_{\ombp}^{}.
\end{equation}
Thus $N^*_\omb$ is the number of observed cases that appear on all the lists in $\omb$, regardless of whether they do or do not appear on other lists.  For example, $N^*_{12}$ is the total number of cases that are on both lists 1 and 2, while $N_{12}^{}$ is the number of individuals that are on lists 1 and 2 but not on lists 3, 4, \ldots \ . We will call $\{i,j\}$ a {\em non-overlapping pair} of lists if $N^*_{ij} = 0$, so that no individual appears on both lists.
The main objective of this paper is to develop estimation procedures and algorithms that properly account for these non-overlapping pairs of lists.

Because of the restriction to $\ombp \ne \emptyset$ in the defining sum, the quantity $N^*_\emptyset$ does not include the dark figure but is the sum of all the observed values $N_\ombp^{}$, the total of individuals actually captured at some point in the study.


\subsection{The Poisson Loglinear Model}

A standard model for the analysis is the Poisson loglinear model as set out by \cite{Cormack1989}.   This assumes that, independently for each $\omb$,
\begin{equation} \label{eq:pardef}
N_\omb^{} \sim \mbox{Poisson}(\mu_\omb^{})
\;\; \mbox{with} \;\;
\log \mu_\omb^{} = \sum_{\thetab \subseteq \omb} \alpha_{\thetab}^{}
\end{equation}
for certain parameters $\alpha_{\thetab}^{}$ indexed by the possible capture histories.  Capture histories are used in two different ways, firstly to index the observed data, and secondly to index the parameters.
Usually, but not invariably,  the letter $\omb$ will be used when observations $N_\omb$ are indexed and $\thetab$ for parameters $\alpha_\thetab$.  The index $\ombp$ will be used in either case, as required.

Thus, for example, the dark figure has expected value $\exp \alpha_\emptyset^{} $, while the expected value of $N_{13}^{}$ is $\exp(\alpha_\emptyset + \alpha_{1}^{} + \alpha_{3}^{} + \alpha_{13}^{})$.  Denoting by $\hat{\alpha}_\emptyset$ the maximum likelihood estimate of the parameter $\alpha_\emptyset$, the estimate of the total population size will be
$N^*_\emptyset+ \exp{\hat{\alpha}_\emptyset}$, the sum of the total number of cases actually observed and the estimate of the dark figure.

Altogether, there are $2_{}^t$ parameters $\alpha_{\thetab}^{}$, corresponding to the $2_{}^t$ capture histories including the null capture history.
There are only $2^t - 1_{}$ observable data points $N_\omb$ from which to estimate the parameters; without placing constraints on the $\alpha_{\thetab}^{}$ parameters, the model is not identifiable.

As \cite{Cormack1989} sets out, the natural approach is to set some of the $\alpha_\thetab$ to zero, and then to estimate the remainder by maximum likelihood; for example, one may set all coefficients indexed by third- or higher-order histories to zero, and we will do this throughout. Even if all the two-list coefficients (those indexed by pairs of lists) are included, the number of parameters to be estimated is $1 + t + \frac{1}{2} t (t-1) \leq 2^t - 1$ provided $t \geq 3$.  Model choice then reduces to deciding which two-list coefficients to include, and will be discussed further in Section~\ref{sec:modelfit}.  For any particular choice of coefficients, the estimation can be put into a standard generalized linear model formulation.

A consequence of the definition is that, for each $\omb$,
$$
N^*_\omb \sim \mbox{Poisson}(\mu^*_\omb) \mbox{~where~} \mu^*_\omb = \sum_{\ombp \supseteq \omb, \ombp \ne \emptyset} \mu_{\ombp}.
$$
Unlike the $N_\omb$, the $N^*_\omb$ are not independent.   For example, if capture histories $\omb$ and $\ombp$ share any lists, then the variables
 $N^*_\omb$ and $N^*_\ombp $ will be dependent.

\subsection{The Log Likelihood Function}

Before considering the treatment of non-overlapping pairs of lists, we derive some properties of the likelihood function.
Let $\Theta$ be the collection of indices of parameters included in the model, and $\alb = ( \alpha_\thetab^{} : \thetab \in \Theta )$ the vector of parameters to be estimated.
Note that $\Theta$ always contains $\emptyset$.
%
%
Up to an additive constant depending only on the data, the log likelihood is given by
$$
\ell ( \alb | {\bf N} ) =
\sum_{\omb \ne \emptyset} \{ N_\omb^{} \log (\mu_\omb^{}) - \mu_\omb^{} \}.
$$
Substituting the definition of the model, reversing the order of summation, and then substituting the definition (\ref{eq:nstardef}), to obtain
\begin{equation}
\sum_{\omb \ne \emptyset}  N_\omb^{} \log (\mu_\omb^{})  = 
\sum_{\omb \ne \emptyset}\{ N_\omb^{} \! \! \! \! \sum_{ \thetab \subseteq \omb, \thetab \in \Theta}
\!\!\! \! \alpha_{\thetab}^{}\}
 = \sum_{ \thetab \in \Theta} \{ \alpha_{\thetab}^{}\! \! \! \!\sum_{\omb \supseteq \thetab, \omb \ne \emptyset}\! \! \! \!N_\omb^{} \}
 = \sum_{ \thetab \in \Theta } \alpha_{\thetab}^{} N^*_{\thetab}.
\label{eq:likterm1}
\end{equation}
%
Turning to the other term in the log likelihood,
$$
- \sum_{\omb \ne \emptyset } \mu_\omb^{} =
\sum_{\omb \ne \emptyset } \left \{- \exp \left [ \sum_{ \thetab \subseteq \omb, \thetab \in \Theta } \alpha_{\thetab}^{}\right ] \right \}  = C(\alb),
$$ 
say.   Regarded as a function of the $\alpha_\thetab^{}$, each $\mu_\omb^{}$ is an increasing function of each of its arguments, and hence $C(\alb)$ is a decreasing function of each of its arguments $\{ \alpha_\thetab^{}: \thetab \in \Theta \}$.  Furthermore, $C(\alb)$ is 
a sum of concave functions of linear combinations of its arguments, so $\ell ( \alb | {\bf N} )$ is the sum of a linear and a concave function, and hence is a concave function.  However, as \cite{Fienberg2012} show in a much more general and abstract context, and as we shall see below, the maximum likelihood estimate of $\alb$ need not be unique or even exist at all.

\label{subsubsec:suffstat}

The expressions for the components of the log likelihood function demonstrate the following, which will be useful in our discussion of model choice:
\begin{enumerate}
\item
The statistics $\{ N^*_\thetab:  \thetab \in \Theta \}$ are jointly sufficient for the parameters $\alb$.
\item
Given any $\omb$ in $\Theta$, $N^*_\omb$ is sufficient for $\alpha_\omb^{}$ if all the other parameters $\{ \alpha_\ombp^{}: \ombp \in \Theta, \ombp \ne \omb \}$ are kept fixed.
\end{enumerate}

%


\subsection{Dealing with Non-Overlapping Pairs}
\label{subsec:nonoverlap}

Suppose that $\{i,j\}$ is a non-overlapping pair, so that $N^*_{ij} = 0$, and  that $\alpha_{ij}^{}$ is one of the parameters in the model being fitted, so that $\{i,j\} \in \Theta$.   In the terminology of \cite{Fienberg2012} we allow an extended maximum likelihood estimate, which means that that the parameters may take values in $[-\infty, \infty)$.  If a parameter $\alpha_\thetab = -\infty$ then we will have $\mu_\omb = 0$ for all $\omb \supseteq \thetab$, so the actual Poisson parameters will still all be finite. This section gives an elementary recapitulation of some of the results \cite{Fienberg2012} cast into our specific framework.

In the first term (\ref{eq:likterm1}) of the log likelihood, the coefficient of $\alpha_{ij}^{}$ is zero, so the maximum likelihood estimate of $\alpha_{ij}^{}$ will be obtained by maximizing $C(\alb)$.
Because $C(\alb)$ is a decreasing function of each of its arguments, whatever the value of the other parameters the likelihood will be maximized as $\alpha_{ij}^{} \rightarrow - \infty$.
The maximum likelihood estimate of $\alpha_{ij}^{}$ may therefore be regarded as $\alpha_{ij}^{} = - \infty$.
This explains why existing software packages yield errors or warnings if there are non-overlapping pairs in the data and the corresponding parameters are in the model.
Because the linear model is expressed in terms of the logarithm of the Poisson parameter, the value $- \infty$ for $\alpha_{ij}^{}$ gives the value zero for $\mu_\omb^{}$ for all $\omb \supseteq \{ i,j\} $, a legitimate value for the actual Poisson parameters, regarding a Poisson distribution with parameter zero to be the degenerate distribution with value zero.

Substituting these zeroes for $\mu_\omb^{}$ back into the expression for the log likelihood yields, writing $\alb^\dagger_{ij}$ for the vector of parameters with $\alpha_{ij}^{}$ excluded,
$$
\ell ( \alb^\dagger_{ij} | {\bf N},\alpha_{ij}^{}=-\infty  )  =
\sum_{\omb \neq \emptyset, \omb \not \supseteq \{ i, j\}} \{ N_\omb^{} \log (\mu_\omb^{}) - \mu_\omb^{} \}.
$$
This is exactly the Poisson log likelihood based on all the observations except those for the $2^{t-2}_{}$ capture histories which include both $i$ and $j$.  Note that the sum is over $\omb$ that do not include the set $\{i,j\}$, in other words both of $i$ and $j$.
If there is more than one non-overlapping pair in $\Theta$, the same calculations can be carried out for each pair, leading to the following algorithm:
\begin{enumerate}
\item Initially define $\Omega^\dag$ be the set of all non-null capture histories and $\Theta^\dag = \Theta$.
\item For each $\{i, j\}$ in $\Theta$ for which $N^*_{ij} = 0$, record that the maximum likelihood estimator of $\alpha_{ij}^{}$ is $-\infty$ and remove $\alpha_{ij}$ from the set of parameters $\Theta^\dag$ yet to be estimated.
\item For each such $\{i, j\}$ also remove from $\Omega^\dag$ all $\omb$ for which $\omb \supseteq \{i, j\}$ (because $N^*_{ij} = 0$ the corresponding $N_\omb$ will all be zero).
\item Use the standard generalized linear model approach to estimate the parameters with indices in $\Theta^\dag$ from the observed counts of the capture histories in $\Omega^\dag$.  The set $\Theta^\dag$ comprises all the two-list parameters in the model that are not estimated to be $-\infty$.   
\end{enumerate}

In the next section, we will see that the final step should also involve an explicit check for the existence and identifiability of the parameter estimates.



\subsection{How Existing Methods go Wrong}
\label{subsec:whatgoeswrong}

Where there is a pair of non-overlapping lists, existing methods typically iterate towards a large negative estimate for the
corresponding parameter $\alpha_{ij}$, only stopping because the number of iterations exceeds a prescribed limit, or because the second derivative of the log likelihood is numerically nearly zero.   An error or warning message may be produced. By contrast, our approach deals explicitly with $\alpha_{ij}$, immediately giving it the value that maximizes the likelihood over the extended range $[-\infty, \infty)$.
Once the
parameters corresponding to non-overlapping pairs of lists have been correctly estimated, all the other parameters are estimated by an iterative process which converges rapidly and does not yield any errors.

Suppose, for the moment, that a large negative value of $\alpha_{ij}$ is used, say $\alpha_{ij} = -20 $ rather than $\alpha_{ij} = -\infty$.
For practical purposes $e^{-20}$ is zero, so the fitted values of $\mu_\omb$ will be essentially zero for all $\omb \supseteq \{i, j\}$ and the corresponding terms will make no contribution to the maximization of the likelihood of the other parameters.  Hence, the fitted values of the other parameters will be much the same as in our approach which actually estimates the parameter $\alpha_{ij}$ correctly.  We are not fitting a different model than other approaches; rather, we are correctly fitting a model which other approaches can only fit approximately and in an unsatisfactory way.

Not only is it inelegant to use an iterative method to approach a known $-\infty$ value of a parameter, but it leads to misleading estimates of the precision of the estimates.  Because the second derivative of the log likelihood also rapidly tends to zero, the estimated parameter tends to have very large reported standard error, suggesting that its estimate is essentially uninformative.  Furthermore, standard packages use approaches to inference and model choice based on likelihood and information criteria.  The asymptotic theory and arguments behind these approaches, for example \cite{Wilks1938} and \cite{Akaike1974}, break down when parameters are at an extreme of their ranges, as is the case in our application for the parameters corresponding to non-overlapping lists (see Section 1 of the supplementary materials for a simulation example illustrating that the likelihood asymptotics do not hold).

An exploration of the possibly misleading precision estimates, for two real data sets, is given in Table \ref{tab:glmnogood}.  The data sets are from the UK \citep{HomeOffice2014} and The Netherlands \citep{DiCrHeKr17}, both tabulated in \cite{Silverman2019}.  In each case the data consist of six lists, and in both cases there are two non-overlapping pairs. We will see that the corresponding parameters are significant in one case but not the other.  The standard errors and $p$-values for {\tt glm} are those produced using the default method for that routine.   

\begin{table}
\caption{Comparison of the performance of standard approaches using {\tt glm} with the method set out in this paper.  The quantity estimated is the parameter $\alpha_{ij}$ for the non-overlapping pair $(i,j)$ under consideration.  The model fitted includes all two-list parameters.
\label{tab:glmnogood}}
\medskip
\centering
\begin{tabular}{||c|c|ccc|cc||}
\hline\hline
  \multirow{2}{*}{Data set}        &
  \multirow{2}{*}{Pair}& \multicolumn{3}{c|}{Standard} & \multicolumn{2}{c||}{Proposed} \\
&  & Estimate & Std err & $p$-value & Estimate & $p$-value \\
 \hline
     \multirow{2}{*}{Netherlands}     &  I:K 	
          & $-20.79	$ 		& 5778 			& 0.997
  				&  $-\infty$ & $9.1 \times 10^{-4}$
  				\\
         & K:R
 				& $-19.96$ & 2783 & 0.994
 				& $-\infty$ 		&  $2.1 \times 10^{-5}$
 				\\
 \hline
 \multirow{2}{*}{UK}				&  LA:GP 		& $-19.08$ & 5350 & 0.997
 				&       $-\infty$ & 0.13 \\
  & LA:NCA &$ -19.19$ & 7968 & 0.998
 			 & $-\infty$ &  0.30\\
 			 \hline \hline
 \end{tabular}
 \end{table}

The table shows the result of fitting the model including all two-list effects, using two algorithms, one being a `standard' approach \citep{Rivest2012} which makes use of the R program {\tt glm}, and the other the method set out above.  In the standard approach, the call to {\tt glm} actually records convergence, but after 21 and 22 iterations respectively, which is close to the default maximum number 25 of iterations in {\tt glm}.  In both cases a warning is generated.   By contrast, the call to {\tt glm} within our approach only requires 6 or 7 iterations.  The estimates of all the other parameters, as expected, are virtually the same in both cases.   The $p$-value for our approach is the probability that  the non-overlap of the relevant pair could have occurred by chance when the model is fitted without the corresponding parameter; see Section \ref{subsec:signif} below.  It can be seen that the effects are highly significant for the Netherlands data but not significant for the UK data.   The reported standard errors and $p$-values are not meaningful for the standard approach.

In fact there are additional aspects not handled by current methods that need to be addressed, even if one allows for the parameters to be estimated over the extended range $[-\infty, \infty)$, and these are discussed in Section \ref{subsec:identifiability}.

\subsection{Existence and Identifiability Issues}
\label{subsec:identifiability}

Two estimability issues may arise when applying multiple systems estimation to sparse data, both of which will mean that the model will not give a well-defined finite estimate of the population size.

One possibility is that there is no value of the parameter vector $\alb$ that maximizes the likelihood, even allowing the extended range $[-\infty, \infty)$ for the parameters.  The other, separate, possibility is that (whether or not the likelihood can be maximized) there is parameter redundancy and the estimates are not identifiable.   We discuss the existence question first.

\cite{Fienberg2012a} show that existence of the estimate can be checked by solving a linear programming problem.
Defining $\Theta^\dag$ and $\Omega^\dag$ as in the algorithm set out in Section \ref{subsec:nonoverlap}, let $\Abm$ be the incidence matrix that maps the parameters in $\Theta^\dag$ to the logarithm of the expected values of the counts of capture histories in $\Omega^\dag$.  From (\ref{eq:pardef}), for $\thetab \in \Theta^\dag$ and $\omb \in \Omega^\dag$,  $\Abm_{ \omb \thetab} = 1$ if $\thetab \subseteq \omb$ and 0 otherwise.  Let $\tbm$ be the vector of sufficient statistics $N^*_\thetab$ for $\thetab \in \Theta^\dag$.
Then set up the linear programming problem of finding the maximum value of $s$
over all scalars $s$ and all real vectors $\xbm = (x_\omb, \omb \in \Omega^\dag)$ satisfying the constraints
\begin{equation}\label{eq:lp}
\Abm^T {\bf x} ={\bf t} \;\;\; \mbox { and } \;\;\;
x_\omb - s \geq 0 \mbox{ for all $\omb \in \Omega^\dag$}.
\end{equation}
A necessary and sufficient condition for a maximum likelihood estimate of $\alb$ to exist (possibly allowing some parameters to be $-\infty$) is that the maximizing value $s_{\rm max}$ of $s$ is strictly greater than 0.

Setting $x_\omb = N_\omb$ for all $\omb$ and $s= \min N_\omb$ will yield a feasible
solution satisfying (\ref{eq:lp}).   Hence $s_{\rm max}$ will be at least the minimum of the observed $N_\omb$ over $\Omega^\dag$. In the non-sparse case, where every combination of capture histories is observed at least once, this minimum will be strictly positive and hence the maximum likelihood estimator will always exist.   

The other possibility is that, even if the likelihood can be maximized, the parameters are non-identifiable, so that the estimate is not unique, a state of affairs also called parameter redundancy; see, for example, \cite{Far2019}. 
The model will be identifiable if and only if $\Abm$ is of full column rank.  We show in Section 3 of the supplementary materials that non-identifiability can only arise if all list pairs are in the model and if the data are so sparse that every set of three lists contains at least one non-overlapping pair.  This condition is easily checked.


\cite{Fienberg2012} point out that most or all standard generalized linear modeling packages fail to check for existence of estimates. Nor do programs necessarily report unidenfiability directly, more often arbitrarily removing one or more of the parameters.  Unless every possible capture history is actually represented in the observed data, therefore, it is important to check that a potential model gives a strictly positive value for the linear programming problem.  If the full model containing all two-list parameters is being fitted then, in addition, identifiability should also be checked.  If the model fails on either count it should be ruled out.  These checks incur only a small computational overhead.

\begin{table}[h]
\caption{\label{tab:demo}An artificial data set with three lists. In this data set, there are no cases with capture histories AC, BC or ABC.}
\centering
\bigskip
\begin{tabular}{||c|c||c|c||}
\hline \hline
\multicolumn{2}{||c||}{Cases observed} & \multicolumn{2}{c||}{Cases observed} \\
\multicolumn{2}{||c||}{only on one list} & \multicolumn{2}{c||}{on exactly 2 lists} \\ \hline \hline
List                   & Number                     & Lists                      & Number                     \\ \hline
A                        & 40                         & A \& B                     & 6                          \\
B                        & 30                         &                           &                            \\
C                        & 20                         &                           &                            \\ \hline \hline
\end{tabular}

\end{table}
A simple example is given in Table \ref{tab:demo}. As there are three possible two-way interactions, there are $2^3=8$ possible choices of the two-list parameters to include in the model.   We summarise the linear programming output $s_{\rm max}$ and test results in Table~\ref{tab:summaryofres}.


\begin{table}[h]
\caption{Summary of linear programming output and test result for all possible choices of two-list effects to include in the model.   For the model containing all three two-list effects, there are finite values of the Poisson means 
$\mu_\omb$ that maximize the likelihood, so $s_{\rm max} > 0$, but these do not correspond to unique values of parameters in the model.  
\label{tab:summaryofres}}
\medskip
\centering
\begin{tabular}{||c|c|c||}
\hline\hline
  \multirow{1}{*}{Two-list parameters included}        &
  \multirow{1}{*}{Test result}& \multicolumn{1}{c||}{$s_{\rm max}$}\\

 \hline
     \multirow{1}{*}{none}     &  no error
          & 1.2	
  \\
 \hline
 \multirow{1}{*}{$\alpha_{AB}$}				&  nonexistent MLE		& 0
 				 \\
\hline
 \multirow{1}{*}{$\alpha_{AC}$}				&  no error		& 3
 				 \\
\hline
 \multirow{1}{*}{$\alpha_{BC}$}				&  no error		& 3
 				 \\
\hline
 \multirow{1}{*}{$\alpha_{AB}, \alpha_{AC}$}				&  nonexistent MLE 		& 0
 				 \\
\hline
 \multirow{1}{*}{$\alpha_{AB}, \alpha_{BC}$}				&  nonexistent MLE 		& 0
 				 \\
\hline
 \multirow{1}{*}{$\alpha_{AC}, \alpha_{BC}$}				&  no error 		& 6
 				 \\
\hline
 \multirow{1}{*}{$\alpha_{AB}, \alpha_{AC}, \alpha_{BC}$}				&  unidentifiable 		& 6
 				 \\

 			 \hline \hline
 \end{tabular}
 \end{table}
 The results show that there is no immediate hierarchical relationship between models that do or do not satisfy the criterion for estimates to exist.  For example, the linear program result is zero for the model including AB and AC, but either adding the third effect BC, or removing AB, will yield a model for which the result is strictly positive.  This issue is elucidated further in
Section \ref{subsubsec:lp} below.




\subsection{Checking All Models}
\label{subsec:checkallmodels}
Given a particular data set, it is useful in certain contexts to check that the estimates exist no matter which two-list terms are included in the model.  An appropriate algorithm allows the Fienberg-Rinaldo conditions to be confirmed much more quickly than the brute force approach of simply checking the criteria for every possible model.   It will be assumed throughout that the model contains the intercept parameter $\alpha_\emptyset$ and the main effect parameters $\alpha_i$ for $i = 1, \ldots, t$.  The model choice to be made is which, if any, of the two-list parameters $\alpha_{ij}$ also to include.  Because there are $\frac{1}{2} t(t-1)$ pairs $\{i,j\}$, the number of possible models is $2^{t(t-1)/2}$, which rapidly becomes very large as the number of lists increases.

\label{subsubsec:lp}

Suppose that $\{i,j\}$ is an overlapping pair of lists, in that $N^*_{ij} > 0$, and that the parameter $\alpha_{ij}$ is in the current parameter set $\Theta$.  Consider the effect of removing this parameter from the model.   Because $\{i,j\}$ is an overlapping pair, this will not change the set $\Omega^\dag$, but because $\{i,j\}$ is removed from $\Theta$ it will also be removed from $\Theta^\dag$ (again defining $\Theta^\dag$ and $\Omega^\dag$ as in Section \ref{subsec:nonoverlap}).
In the linear programming problem (\ref{eq:lp}), this will remove one column from the matrix $\Abm$ and the corresponding element of ${\bf t}$.   Hence one constraint will be removed, and therefore the maximum value of $s$ cannot decrease.   Therefore, if the estimate exists for parameter set $\Theta$ it will necessarily exist for subsets of $\Theta$ obtained by removing overlapping pairs.    It follows that, to confirm whether all models satisfy the conditions for estimates to exist, it is only necessary initially to test parameter sets $\Theta$ that include all overlapping pairs, together with a subset (possibly empty) of the non-overlapping pairs.   If there are $M$ non-overlapping pairs in the data, then the number of such models is $2^M$; solving the linear programming problem for all these models is now feasible for a much larger range of data sets than if all $2^{t(t-1)/2}$ models have to be considered explicitly.  If the estimates exist for all such models, then they will exist for all models.   

These checks were carried out for all the data sets discussed in this paper.  For the full UK and Netherlands data, the number of models to be checked by solving a linear programming problem is reduced by a factor of $8192$.   Details for two other data sets are given in Sections~\ref{subsec:neworl} and \ref{subsec:western}.  In every case, in contrast to the example set out in Table~\ref{tab:demo}, the extended maximum likelihood estimate exists and is unique for every possible choice of model.  

In the event that there are models for which the estimate does not exist, the approach can be extended to find a list of all such models efficiently.
Let $\Theta_1$ be the set of parameter indices corresponding to the empty capture history and all capture histories of order 1, $\Theta_2^{\rm non}$ those corresponding to non-overlapping pairs and $\Theta_2^{\rm over}$ overlapping pairs.  The initial search is over all models containing both $\Theta_1$ and $\Theta_2^{\rm over}$.   Suppose it yields a subset $\tilde{\Theta}_2^{\rm non}$ with the property that there is no maximum likelihood estimate within the model with parameter set
$ \Theta_1  \cup \Theta_2^{\rm over} \cup \tilde{\Theta}_2^{\rm non}$.  We then perform a hierarchical search, retaining  $\Theta_1 \cup \tilde{\Theta}_2^{\rm non}$, over models where overlapping pairs are removed.  At the first stage, parameters in $\Theta_2^{\rm over}$ are removed individually and each resulting model checked.  If any such model yields a zero result in the linear program, that is recorded, and the possibility of removing a second overlapping pair is investigated, and so on.  At each stage, if the linear program yields a positive result so that the estimate exists, there is no need to investigate that branch of the hierarchy any further.



\section{Inference and Model Choice}
\label{sec:modelfit}

We now consider how to assess the significance of any particular two-list parameter, and develop a forward stepwise approach to model choice. We also develop the bootstrap procedure for evaluating confidence intervals, and present simulation results comparing the bootstrap with an approach that carries out inference conditional on the model actually selected.

\subsection{Calculating Significance}
\label{subsec:signif}

Given any model defined by parameter set $\Theta$, for any $\omb$ define
$$
\hat{\mu}_\omb^{}[\Theta] = \exp \left( \sum_{\thetab \subseteq \omb, \thetab \in \Theta} \hat{\alpha}_\thetab^{} \right)
$$
where the $\hat{\alpha}_\thetab^{}$ are the maximum likelihood estimates of the $\alpha_\thetab^{}$.  Further, define
$$
\hat{\mu}^*_\omb[\Theta] = \sum_{\ombp \supseteq \omb, \ombp \ne \emptyset} \hat{\mu}_{\ombp}^{}[\Theta].
$$
Under these definitions, $\hat{\mu}_\omb^{}[\Theta]$ and $\hat{\mu}^*_\omb[\Theta]$ are the estimated expected value of $N_\omb$ and $N^*_\omb$ respectively.  The notation $[\Theta]$ makes explicit the dependence on the parameter set $\Theta$.

First, consider how to deal with non-overlapping pairs within the data. 
Suppose that for some $\thetab \in \Theta$ that $N^*_\thetab = 0$.  Should we actually include $\thetab$ in the model?  We test the null hypothesis that $\alpha_\thetab^{} = 0$, which is equivalent to saying that $\thetab$ is not in the model.  We fit the model without $\thetab$ and then consider the $p$-value of a test statistic.  A natural test statistic is
$N^*_\thetab$, because of the results on sufficient statistics in Section \ref{subsubsec:suffstat}. Recall that this is also a Poisson random variable since it is the sum of independent Poisson random variables (see \ref{eq:nstardef} and \ref{eq:pardef}). Hence, we test whether 0 is a surprising value to observe for a Poisson distribution estimated from the data but leaving out the parameter indexed by $\thetab$. If $\thetab$ is in the model, then the observed value has probability one if $\thetab$ takes its estimated value.   

Hence, proceed as follows:
\begin{enumerate}
\item  Fit the model leaving out the parameter $\alpha_\thetab^{}$, in other words using just the parameter set $\Thetabm$.
  For the resulting fitted model, find the estimate $\mut$.
\item
The estimated parameter has $p$-value
$\exp ( - \mut ) $.  This is the estimated probability that $N^*_\thetab = 0$ in the model defined by $\Thetabm$.
\end{enumerate}

Unless we have already checked that the parameter set $\Thetabm$ passes the linear program test for the existence of the maximum likelihood estimate, that should be done; if the model fails that test then the effective $p$-value is zero because the parameter $\alpha_\thetab^{}$ cannot be removed from the model.

This approach can be generalized to construct a (one-sided) $p$-value for {\em any} parameter $\thetab \in \Theta$ whether or not $N^*_\thetab = 0$.   
The $p$-value
is the minimum of $F_{\rm Poiss}( N^*_\thetab, \mut)$ and $\tilde{F}_{\rm Poiss}( N^*_\thetab, \mut)$.  Here
$F_{\rm Poiss}( n, \lambda)$ is the lower tail probability that a $\rm{Poiss}(\lambda)$ random variable $X$ satisfies $X \le n$, while $\tilde{F}_{\rm Poiss}(n, \lambda)$ is the probability that $X \ge n$.

An alternative approach is to use the sufficient statistic $N^*_\thetab$ for $\alpha_\thetab^{}$ evaluated against its distribution conditional on the observed values of the sufficient statistics in the model with parameters indexed by $\Thetabm$, rather than, as we have, against its unconditional distribution on the estimated model. The conditional distribution does not seem to be easily tractable, but this is an interesting avenue for future research.





\subsection{Model Fitting}
\label{subsec:modelfitting}
The model-fitting procedure is detailed stepwise, as follows:
\begin{itemize}
\item Step 1: Set a threshold value for the $p$-value and fit the model with the main effects parameters only.
\item Step 2: Consider in turn each two-list parameter not already added to the model, and check that adding it to the model would not lead to a nonexistent estimate (or to non-identifiability if the full two-way model is proposed).
\item Step 3: Among those parameters that pass the checks, find the one with the smallest $p$-value, using the approach set out in Section \ref{subsec:signif}.
If that $p$-value is less than or equal to the given threshold, add the parameter to the model, and go back to Step 2. If the $p$-value is greater than the threshold, finish.
\end{itemize}

Note that in Step 2 all two-list parameters not already included are considered, whether the pairs they correspond to are overlapping or non-overlapping. The method is akin to forward stepwise regression.  Note also that if the algorithm set out in Section \ref{subsec:checkallmodels} has already demonstrated that nonexistence and non-identifiability cannot arise for any model for the data set in question, then the check in Step 2 is not necessary. 

It remains to choose the threshold $p$-value.
We conduct a detailed simulation study in Section \ref{subsec:threshsim} which points to the choice $p=0.02$, and that is the value which we would suggest, but users might wish to explore the sensitivity of the result to adjusting the parameter.  

\subsection{Boostrapping to Find Confidence Intervals}
\label{subsec:bootstrap}
In general, current approaches find confidence intervals for the population size conditional on the terms actually included in the model, either for the Poisson log-linear model itself, or for modifications such as the multinomial model considered by \cite{San1984}.   Because the choice of model itself depends on the observed data, it is preferable to construct confidence intervals which take account directly of the effect of model selection.  
A natural way of doing this is to use a bootstrap approach, which will also take account of any biases that the model selection approach may introduce.  
The $BC_a$ methodology of \cite{DiCEfr96} gives second-order accuracy and does not depend on any transformation of the scale on which the data are observed and the estimate of the total population made.

The observed data in our case are the original $m$ observed capture histories $(\xib_1, \xib_2, \ldots, \xib_m)$.    To construct each bootstrap sample, we could draw a random sample $(\xib_1^{\rm boot}, \xib_2^{\rm boot}, \ldots, \xib_m^{\rm boot})$ of size $m$, with replacement,
from the original data.
If we denote by $N_\omb^{\rm boot}$ the number of times the capture history $\omb$ occurs in the bootstrap sample, then the $N_\omb^{\rm boot}$ have a multinomial distribution corresponding to $m$ trials and probabilities proportional to the original $N_\omb$.
In practice, therefore, the $\xib_i^{\rm boot}$ are not actually constructed, but we sample direct from the multinomial distribution of the capture history totals.    The parameter for the number of trials in the multinomial distribution is the number $m$ of capture histories actually observed and does not depend on any estimate of the dark figure.

For each bootstrap sample, we carry out the stepwise fitting procedure and obtain an estimate (bootstrap replication) of the population size.   There is no constraint on choosing the same model.  The $BC_a$ confidence intervals use percentiles of the bootstrap distribution of the population size,
but they adjust the percentile actually used.   The adjusted percentiles depend on an estimated bias parameter $\hat{z}_0$, defined so that $ \Phi(\hat{z}_0)$ is the proportion of the bootstrap estimates that fall below the estimate from the original data, and an estimated acceleration factor $\hat{a}$, whose derivation depends on
a jackknife approach.

The jackknife requires the population size to be estimated from every sample constructed from the original data by leaving out one of the data points $\xib_i$.   However, the number of jackknife estimates that need to be evaluated can usually be dramatically reduced, making for considerable computational savings, because the number of distinct values taken by the $\xib_i$, the number of different capture
histories actually observed, is in general much smaller than $m$.
If there are $K$ capture histories for which $N_\omb > 0$, only $K$ jackknife estimates actually have to be calculated.  These are then weighted in the calculations by the number of times $N_{\omb}$ that the value
$\omb$ appears in the original sample.   To be explicit, let $\hat{\theta}_{(i)}$ be the estimate of the population size constructed from the original sample leaving out capture history $\xib_i$.  The effect of leaving out that capture history is to reduce $N_{\xib_i}$ by one, and so
$\hat{\theta}_{(i)} = \hat{\theta}_{\xib_i}^{(-1)},$ where, for each capture history $\omb$ actually observed in the data, $\hat{\theta}_\omb^{(-1)}$ is the estimate of the population size from the original sample but with $N_\omb$ replaced by $N_\omb -1$.
Only the $K$ values $\hat{\theta}_\omb^{(-1)}$ have to be calculated.

To calculate the acceleration factor, let $\hat{\theta}_{(\cdot)}$ be the average of the jackknife estimates $\hat{\theta}_{(i)}$.
Then
$$
\hat{\theta}_{(\cdot)} = m^{-1} \sum_{\omb} \sum_{i: \xib_i = \omb}  \hat{\theta}_{(i)}  = m^{-1} \sum_{\omb: N_\omb>0} N_\omb \hat{\theta}_\omb^{(-1)}.
$$
Applying a similar weighting argument to the defining equations (6.6) and (6.7) of \cite{DiCEfr96}, the estimated acceleration factor $\hat{a}$ is then given by
$$
\hat{a} =\tfrac{1}{6}
	\left \{ \sum_{\omb: N_\omb>0} N_\omb ( \hat{\theta}_{(\cdot)} - \hat{\theta}_\omb^{(-1)} )^3 \right \}
            \left \{ \sum_{\omb: N_\omb>0} N_\omb ( \hat{\theta}_{(\cdot)} - \hat{\theta}_\omb^{(-1)})^2 \right \}^{-3/2}  .
$$

These values of the parameters $\hat{z}_0$ and $\hat{a}$ are then used to choose the appropriate percentiles of the bootstrap distribution, using the standard $BC_a$ formulation.

\subsection{Some Simulation Results}
\label{subsec:bootandbicsim}
In order to compare our method with the standard BIC approach as implemented within {\bf Rcapture} \citep{Rivest2012}, a simulation study was carried out.  The model fitted to the five-list UK data by the stepwise approach was used as a starting point.  For this model, the predicted probabilities of each of the 32 possible capture histories (including the empty capture history) were calculated.  The overall population size was that estimated by the model fit.   The reason for using this model as a basis for a simulation is that it is reasonable to suppose that it will display features likely to be seen when using the methods in the human trafficking context.  An example with five lists was used so that the repeated use of the BIC method does not become computationally burdensome.   

The population size and the capture history probabilities were regarded as fixed, and were used as parameters for multinomial sampling to create 500 simulated data sets.  For each simulation, population estimates and confidence intervals were constructed both using the BIC approach and using the stepwise method we have set out.   For the BIC method, multinomial confidence intervals using the routine {\tt closedpCI.t} within {\bf Rcapture} were found; the confidence intevals for the stepwise method were constructed using the $BC_a$ approach.
Because the simulations are constructed from a model with known population size, it was possible to assess the accuracy of the estimation.   The root mean square error of the estimation was 3057 for the stepwise approach and 5834 for the BIC method.   The root mean square errors of the estimate of the log of the population size were 0.19 and 0.34 respectively, so again, for this example, the stepwise approach has much better performance.

The coverage rate of the estimated confidence intervals was also determined.  For the stepwise method using the BCa approach, the nominal 95\% confidence interval contained the true value for 90\% of the simulations, while the nominal 80\% intervals had an actual coverage rate of about 70\% (346 out of the 500 replications).   While these rates are not perfect, the corresponding observed coverage rates for the methods using routines in Rcapture were considerably lower, 61.4\% and 42.8\% respectively.

\section{Empirical Applications}
\label{sec:applications}

In this section, our methods are applied to two data sets relating to victims of modern slavery and human trafficking in the USA.  Both data sets display the sparseness of overlapping entries typical of data collected in this field.   In addition they are also typical of data collected in local regions (rather than entire large countries) in having relatively small counts, with the total number of observed cases in the hundreds and not the thousands.  The two data sets, together with those discussed in Section \ref{subsec:whatgoeswrong} are then used to construct a simulation study investigating the appropriate choice of threshold parameter. 
\subsection{The New Orleans Data}
\label{subsec:neworl}
\cite{Bales2018} discuss a data set collated from a number of sources in New Orleans, given in Table \ref{tab:NewOrl}.
\begin{table}[H]
\caption{\label{tab:NewOrl}Victims related to modern slavery and trafficking in New Orleans. Numbers of cases on each possible combination of lists, leaving out combinations for which no cases were observed.  For reasons of confidentiality the lists are labelled uninformatively.}
\medskip
\centering
\begin{tabular}{||c|c||c|c||c|c||}
\hline
\multicolumn{2}{||c||}{Cases observed} & \multicolumn{2}{c||}{Cases observed} & \multicolumn{2}{c||}{Cases observed} \\
\multicolumn{2}{||c||}{only on one list} & \multicolumn{2}{c||}{on exactly 2 lists} & \multicolumn{2}{c||}{on exactly 3 lists} \\  \hline \hline
List                  & Number                  & Lists                      & Number                      & Lists                      & Number                    \\ \hline
A                      & 25                      & A\&C                       & 1                           & A\&C\&G                    & 1                         \\
B                      & 5                       & A\&D                       & 2                           & A\&D\&E                    & 1                         \\
C                      & 70                      & A\&E                       & 1                           &                            &                           \\
D                      & 33                      & B\&F                       & 1                           &                            &                           \\
E                      & 6                       & C\&D                       & 1                           &                            &                           \\
F                      & 6                       & C\&E                       & 1                           &                            &                           \\
G                      & 6                       & C\&G                       & 1                           &                            &                           \\
H                      & 21                      & D\&E                       & 2                           &                            &                           \\
                       &                         & E\&H                       & 1                           &                            &                           \\ \hline \hline
\end{tabular}

\end{table}

Altogether there are eight lists, and so the full incidence table of observable capture histories, including those combinations for which the actual observed number is zero, has 255 rows.  The null capture history, corresponding to the dark figure, of course cannot be observed, and estimating it is the task of the analysis.
There are 28 possible pairs of lists, and of these there are 18 non-overlapping pairs.
Using the threshold $p=0.02$ fits a model including one two-list parameter, indexed by the pair DE.   The point estimate of the total population size is 1184.  
The $BC_a$ bootstrap confidence interval, based on 1000 bootstrap replications, is (717, 1657).
If main effects only are chosen (which will be the case for the threshold $p=0.01$ or smaller), then
the resulting model yields a  95\% confidence interval of (644, 1618) with a point estimate of 997.  Arguably, with as many as 28 possible two-list parameters, there is some merit in using a smaller threshold.

Because some of the list counts are so small, the effect of combining the four smallest lists into one, to give a five-list version of the data, was also investigated.  If this is done, none of the two-list parameters is significant even at the 5\% level, and the $BC_a$ confidence interval is (589, 1703) with a point estimate of 1034, a result very close to that yielded by the eight-list data with the smaller threshold.   As a further illustration of the issues discussed earlier in the paper, and the need to handle non-overlapping lists in the way we have developed, the {\bf Rcapture} routine {\tt closedpMS.t} was used to fit every possible choice of model with two-list effects.  There are 1024 such models, and in only 124 of these was the fit successful without generating a warning. In the majority of cases there was a warning that the asymptotic bias is large.

Return to the full data as an example for the methodology set out in Section \ref{subsec:checkallmodels}.  There are $2^{28}$ possible models, and 18 non-overlapping pairs.    To check every possible model for existence of the maximum likelihood estimate, there are $2^{18}$ linear programming problems to solve.  This check, which would have been impossible if all $2^{28}$ models had to be considered explicitly, only takes a few minutes on a standard PC.  Neither of the problems identified in Section \ref{subsec:checkallmodels} arises for any model for these data.

\subsection{The Western Site Data}
\label{subsec:western}

One of two data sets considered by \citep{Farrell2019} is collated from a number of sources in the Western site of a research study in the USA.  The data are given in Table \ref{tab:Western}.

\begin{table}[H]
\caption{\label{tab:Western}Victims related to human trafficking in the Western site of a research study in the USA. Numbers of cases on each possible combination of lists, leaving out combinations for which no cases were observed.  For reasons of confidentiality the lists are labelled uninformatively. }
\medskip
\centering
\begin{tabular}{||c|c||c|c||c|c||}
\hline
\multicolumn{2}{||c||}{Cases observed} & \multicolumn{2}{c||}{Cases observed} & \multicolumn{2}{c||}{Cases observed} \\
\multicolumn{2}{||c||}{only on one list} & \multicolumn{2}{c||}{on exactly 2 lists} & \multicolumn{2}{c||}{on exactly 3 lists} \\  \hline \hline
List                  & Number                  & Lists                      & Number                      & Lists                      & Number                    \\ \hline
A                      & 52                      & A\&C                       & 4                           & A\&C\&E                    & 1                         \\
B                      & 90                      & A\&D                       & 2                           & B\&C\&D                    & 1                         \\
C                      & 114                     & A\&E                       & 5                           &                            &                           \\
D                      & 45                      & B\&C                       & 6                           &                            &                           \\
E                      & 21                      & B\&D                       & 1                           &                            &                           \\
                       &                         & D\&E                       & 3                           &                            &                           \\ \hline \hline
\end{tabular}


\end{table}

Altogether there are 5 lists, and so the full incidence table including those combinations for which the observed number is zero has 31 rows. There are 10 possible pairs of lists, and of these there are 2 non-overlapping pairs.  It is very quick to check that all possible models lead to estimates which exist and are identifiable.

The threshold of $p=0.02$ yields a model including the two-list effect AE, 
with a point estimate of 2483.   The $BC_a$ confidence interval is (1293, 3670).

It should also be noted that the application of {\tt closedpMS.t} to this data set again generated warnings in more than half of the 1024 possible models.  In both this data set and the New Orleans five-list data set, warnings were generated among the top ten models according to the BIC that {\tt closedpMS.t} displays by default, but not, as it happens, by the very top model.    For the Netherlands data considered earlier, six out of the ten top models generates a warning.  


\subsection{Choosing the Threshold:  A Simulation Study}
\label{subsec:threshsim}

In order to gain insight into the appropriate choice of threshold, a simulation study was carried out.  In order to make this relevant to the context of human trafficking, the models considered are all based on the data sets referenced in this paper, in an attempt to ensure that the simulation study is based on data sets which have the kinds of characteristics likely to be encountered.  The data sets considered were the UK, Netherlands, New Orleans and Western site data; in the case of the UK, Netherlands and New Orleans data, both the full and the five-list versions were included, giving seven data sets in all. For each of these, four different models were fitted; the `full' model with all two-list effects included, the model based on main effects only, and the models chosen by the method we set out, using thresholds 0.001 and 0.05, to give a more parsimonious and a less restrictive fit.  In every case, the model fit gives an estimate of the total population and of the probabilities of all possible capture histories.

For each of these 28 test cases, 1000 realizations of the capture history totals were simulated, by drawing from a multinomial distribution with the given population size and capture history probabilities.  Each realization can be conceptualized as an example of a multiple systems sample from a population of known size, with characteristics similar to those likely to be observed in the human trafficking context.
For each realization, estimates of the total population were obtained using a range of thresholds (0.001, 0.002, 0.005, 0.01, 0.02, 0.05, 0.1), as well as for the main effects model and the model with all two-list effects included, corresponding to thresholds 0 and 1 respectively.    The estimates of the total population size were, as one would expect given the log-linear nature of the modeling,  asymmetrically distributed around the true value for each simulation scenario, and a log transformation is appropriate.  With this in mind the measure of accuracy used for each of the 28 sets of simulations, for each threshold parameter, was the mean square error of the logarithm of the estimate of the dark figure.

The general level of mean square error varied quite substantially across the 28 models considered.  In order to take account of this variation, for each threshold, the mean, over the 28 models, of the logarithm of the mean square error was calculated to give an overall score for that threshold.  The threshold with the minimum score is  $p=0.02$.   Further details of the simulation study are given in Section 5 of the supplementary materials to this paper.

\section{Concluding Remarks}

The R software package \SparseMSE \SparseMSEcitep\ includes implementations of all the methodology described in this paper.   In particular, it contains programs to check whether a particular model leads to either of the estimability issues set out in Section \ref{subsec:identifiability} above, and it incorporates these checks within a routine to fit any particular model, or to make the model choice using the stepwise procedure described in Section \ref{subsec:modelfitting}.  It also allows for the possibility of checking all possible models using the approaches discussed in Section \ref{subsec:checkallmodels}. Full details are given in the package documentation.

To conclude, in this paper we have investigated inference for multiple systems estimation using Poisson log-linear models, taking proper account of the possibility that the underlying data tables contain non-overlapping lists, as commonly arises when the data are collected in the context of studies on modern slavery and human trafficking.  We have also set out an approach to model choice and demonstrated the utility and practicality of our approach on real data sets.  This area is especially challenging for methodological development because there is no ``ground truth'' against which methods can be assessed, and frequently there are no details of the data available beyond anonymized list data of the form presented in the tables above. Nevertheless, reliable and stable methods are important for applications in public policy, even if they are conditional on assumptions that it may not be possible to verify.

For simplicity and clarity, the procedure has been discussed and detailed in full for models which consider up to terms indexed by pairs of lists.  In principle, the model fitting and inference aspects can easily be extended to consider models based on higher-order terms, though it seems unlikely that any data sets collected in the contexts of human trafficking would merit this.  For example, if a three-list parameter $\alpha_{123}$ were a candidate for inclusion within the model, then the estimate of $\alpha_{123}$ would be $-\infty$ if the three-list overlap $N_{123}^*$ were empty, and to fit the other parameters one would then remove all capture histories including all three lists 1, 2 and 3 from the {\tt glm} stage.

Similarly, another possible extension is to the case where there is covariate information rather than just presence/absence on various lists.   As in our main discussion suppose there is a pair (or larger set) of lists whose interaction parameter is in the model but for which no overlapping cases are observed for any value of a covariate.   Then the right approach  (depending on the exact details of the modeling) would be to set the corresponding interaction parameter to $-\infty$ and then remove various zero cells containing the non-overlapping set of lists from the fitting procedure for the other parameters including those relevant to covariates.

One possible topic for future research is the combination of our insights with those of \cite{WGDMF1995}, which explores the effect of heterogeneity.  Some of the approaches suggested in that paper may not be available.  For example in the human trafficking context we may not be able to stratify the population, nor may the statisticians analyzing the data have any information about the lists themselves to evaluate the possibility of heterogeneity.  On the other hand, if one is in a position to implement the proposals, then the possibility of effects of the kind we have explored has to be taken into account.



\bibliographystyle{biom}
\bibliography{SparseMSE}
\appendix
\section{Supplementary Materials}
The supplementary materials contain additional details to complement the main paper ``Multiple Systems Estimation for Sparse Capture Data: Inferential Challenges when there are Non-Overlapping Lists". In particular, the document provides:  additional information and details for the simulation carried out to investigate the asymptotic likelihood theory that supports Section 2.5 of the main paper;   the UK and Netherlands data sets, used as empirical examples; justification of the conditions for non-identifiability as mentioned in Section 2.6;  and full details, including R code, of the simulation study conducted to inform a suitable choice of $p$-value threshold, as discussed in Section 4.3.

\subsection{Investigation of the Asymptotic Likelihood Theory}

One of the standard approaches in the literature when fitting log-linear models is to use methods based on likelihood ratios for inference and on information criteria for model choice.
The asymptotic theory and arguments behind these approaches, for example \cite{Wilks1938} and \cite{Akaike1974}, break down when parameters are at an extreme of their ranges, as is the case in our application for the parameters corresponding to non-overlapping lists.  The simulation study to demonstrate this break down is based on
models with three lists and an expected true population
size of 1000, with captures on the various lists being independent, with probabilities $(p_1, p_2, p_3)$, so that the probability of capture history $\omb$ is $\prod_{i \in \omb} p_i \prod_{i \not \in \omb} (1-p_i)$.

The first model used has capture probabilities set to 0.3 for each capture occasion, which is more reminiscent of a classical
mark-recapture study, while the second has capture probabilities set to 0.01, 0.04, and 0.2, which is somewhat typical of the sparse capture case.   Therefore for the first model, if the history contains $k$ captures then the probability is $0.3^k 0.7^{3-k}$. For the second model the probability depends on the actual lists; for example the probability of capture history $\{1,3\}$ is $0.01 \times (1- 0.04) \times 0.2$.

\begin{figure}[H]
\caption{QQ plots based on 10,000 simulations for sparse model (solid line) and for classic model (dashed line) against quantiles of the $\chi^2_1$ distribution. The dotted line $x=y$ is followed
closely by the QQ plot for the classic model. }
\centering
\includegraphics[width=0.5\linewidth]{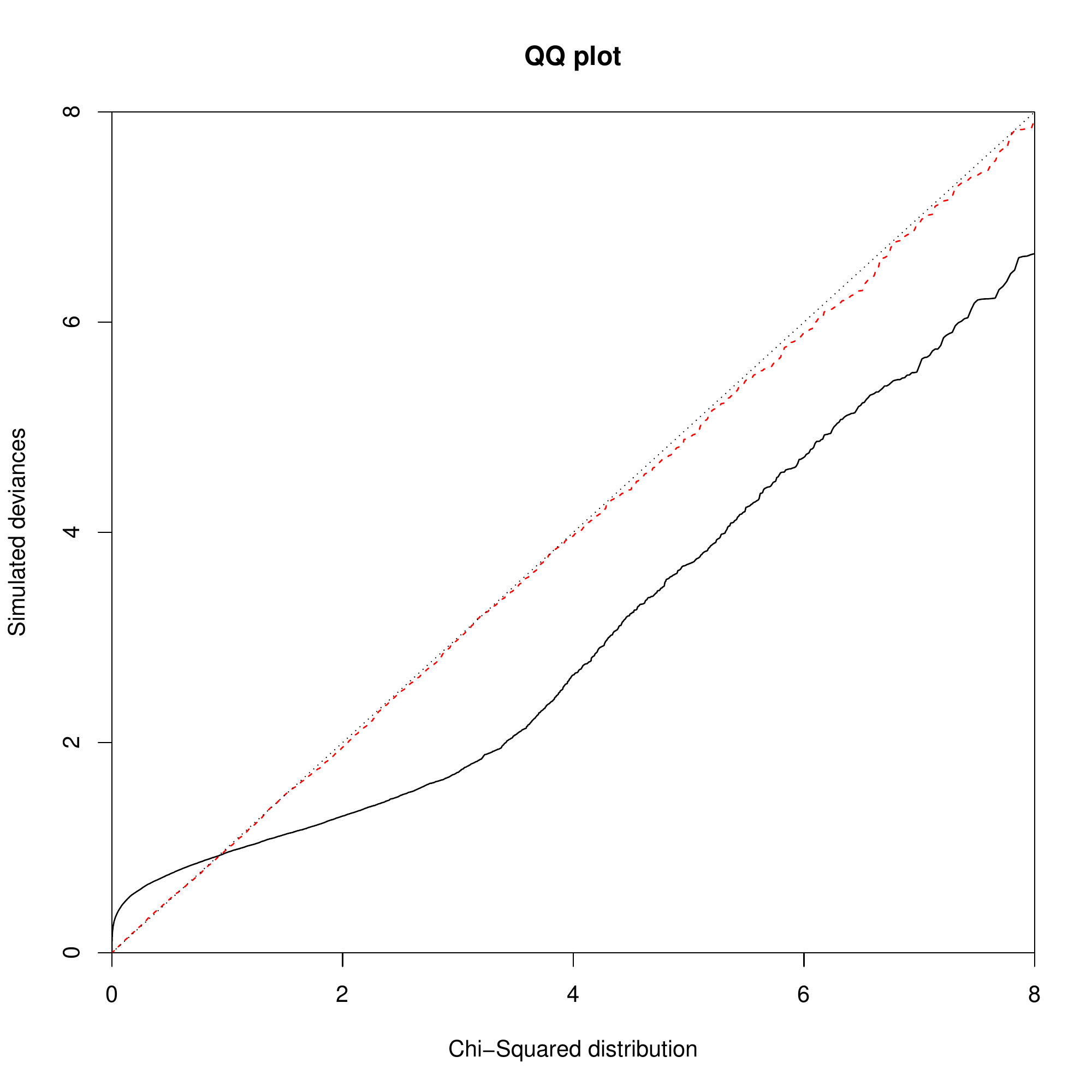}
\label{fig:qqplot}
\end{figure}

The simulation presented is carried out using the Poisson model.  The probabilities are multiplied by 1000 and, for each simulation replicate, Poisson random values with expectations equal to these values are generated to give a full set of observed capture histories; together with the null capture history the expected number of counts (population size) is equal to 1000. The correct model for these data includes main effects only. Inference was carried out both for that model, and for the model with the addition of a two-list parameter indexed by the first two lists. The reduction in deviance between the two models was determined.
In line with the standard asymptotic theory, the QQ-plots presented in Figure \ref{fig:qqplot} show that the $\chi^2_1$ distribution fits the observed deviances well for the ``classic'' model. However, not at all surprisingly, the fit for the sparse model is not good.   This illustrates that likelihood ratio tests, and hence methods based on information criteria, cannot be relied on for fitting models in the sparse context.  

\subsection{Description of Empirical Data Sets}

In Sections 2.5 and 4.3 of the main paper, the UK and Netherlands data sets are used in an empirical application and simulation study. These data sets have both been published previously, but for convenience they are included here.

\subsubsection{The UK Data Set}
As part of the UK Government's Modern Slavery Strategy \citep{HomeOffice2014}, \cite{Sil14} discussed and analyzed a data set collated from a number of sources in the UK.
The lists and corresponding abbreviations are given in Table~\ref{UK_abb}.

\begin{table}[H]
\caption{List names and abbreviations for the UK data set \citep{Sil14}.}
\centering
\begin{tabular}{||l|l||}
\hline\hline
LA & Local authorities \\
NG & Non-government organisations such as charities \\
PF  & Police forces \\
GO & Government organisations \\
GP & The general public, through various routes \\
NCA & The National Crime Agency \\
\hline \hline
\end{tabular}
\\
\label{UK_abb}
\end{table}
In the original analysis, the lists PF and NCA were combined into a single list, to yield what is referred to as the five-list UK data set. The six-list version of the data set was published in \cite{Bales2015}. A rationale for combining the lists PF and NCA is that the National Crime Agency has many of the characteristics of a police force.

Counts of the capture histories over combinations of the lists are given in Table~\ref{tab:UK}. The full incidence table, including those combinations for which the observed number is zero, has 63 rows. There are 15 possible pairs of lists, and of these there are 2 non-overlapping pairs.

\begin{table}[H]
\caption{Potential victims of trafficking in the UK, 2013, as reported to the National Crime Agency Strategic Assessment. From \cite{Bales2015}.}
\medskip
\centering
\resizebox{\columnwidth }{!}{
\begin{tabular}{||c|c||c|c||c|c||c|c||}
\hline
\multicolumn{2}{||c||}{Cases observed} & \multicolumn{2}{c||}{Cases observed} & \multicolumn{2}{c||}{Cases observed} & \multicolumn{2}{c||}{Cases observed} \\
\multicolumn{2}{||c||}{only on one list} & \multicolumn{2}{c||}{on exactly 2 lists} & \multicolumn{2}{c||}{on exactly 3 lists} &\multicolumn{2}{c||}{on exactly 4 lists} \\  \hline \hline
List                  & Number                  & Lists                      & Number                      & Lists                      & Number           & Lists                      & Number         \\ \hline
LA                      & 54                      & LA\&NG                      & 15                           & LA\&NG\&PF                    & 1              &LA\&NG\&PF\&GO   &1           \\
NG                      & 463                       & LA\&PF                      & 19                           & LA\&NG\&GO                    & 1                  &          &       \\
PF                     & 907                     & LA\&GO                       & 3                           &NG\&PF\&GO           &4                 &            &               \\
GO                     & 695                      & NG\&PF                       & 56                           &  NG\&PF\&NCA          &3                 &         &                  \\
GP                    & 316                       & NG\&GO                       & 19                           &  PF\&GO\&NCA         &1                 &          &                 \\
NCA                      & 57                       & NG\&GP                       & 1                           &                            &                       &   &     \\
&                      &   NG\&NCA                       & 3                      &     &                            &                         &  \\
&                      & PF\&GO                      & 69                       &                            &                            &                     &     \\
&                       &PF\&GP                        & 10                       &   &    & & \\                        &                        &PF\&NCA                    &31 &&&& \\
&                    &GO\&GP & 8  &&&& \\
&           &GO\&NCA &6 &&&&                  \\
& &GP\&NCA &1 &&&&  \\ \hline \hline
\end{tabular}}
\label{tab:UK}
\end{table}

\subsubsection{The Netherlands Data Set}

\cite{Cruyff2017} and~\cite{ DiCrHeKr17} analyze a data set collated from a number of sources in the Netherlands. The lists and corresponding abbreviations are given in Table~\ref{tab:Ned_abb}.

\begin{table}[H]
\caption{List names and abbreviations for the Netherlands data set \citep{Cruyff2017}.}
\centering
\begin{tabular}{||l|l||}
\hline\hline
I & Inspectorate SZW\\
K & Border Force \\
O  &Residential centers/shelters\\
P & National Police \\
R & Regional Coordinators \\
Z & Others\\
\hline\hline
\end{tabular}
\\
\label{tab:Ned_abb}
\end{table}

Counts of the capture histories over combinations of the lists are given in Table~\ref{tab:Ned}. The full incidence table, including those combinations for which the observed number is zero, has 63 rows. There are 15 possible pairs of lists, and of these there are 2 non-overlapping pairs. Analysis is carried out both for the full data set, and for the `five-list Netherlands data set' obtained by combining the two smallest lists I and O into a single list.

\begin{table}[H]
\caption{Victims of trafficking in the Netherlands, from~\cite{Cruyff2017}.}
\medskip
\centering
\begin{tabular}{||c|c||c|c||c|c||}
\hline
\multicolumn{2}{||c||}{Cases observed} & \multicolumn{2}{c||}{Cases observed} & \multicolumn{2}{c||}{Cases observed} \\
\multicolumn{2}{||c||}{only on one list} & \multicolumn{2}{c||}{on exactly 2 lists} & \multicolumn{2}{c||}{on exactly 3 lists} \\  \hline \hline
List                  & Number                  & Lists                      & Number                      & Lists                      & Number                    \\ \hline
I                      & 352                      & I\&O                       & 1                           & I\&O\&P                    & 4                         \\
K                      & 1299                       & I\&P                       & 18                           & I\&P\&Z                    & 4                         \\
O                      & 403                      & I\&R                       & 3                           & O\&P\&R                           &2                           \\
P                      & 4466                      & I\&Z                       & 16                          &O\&P\&Z                            &7                           \\
R                      & 650                       & K\&O                       & 1                           &P\&R\&Z                            &1                           \\
Z                      & 632                       & K\&P                       & 44                           &                            &                           \\
                      &                        & K\&Z                       & 4                           &                            &                           \\
                     &                       & O\&P                       & 59                           &                            &                           \\
                       &                         & O\&R                       & 2                           &                            &
\\
                       &                       &O\&Z                         &57                           &                                  &
\\
                       &                       &P\&R                        &82                           &                                   &
\\
                       &                          &P\&Z                       &125                        &                                       &
\\
                       &                          &R\&Z                        &2                         &                                        &
\\ \hline \hline
\end{tabular}
\label{tab:Ned}
\end{table}
\subsection{Non-identifiability}

In Section 2.6 of the main paper we stated conditions for non-identifiability which can be summarised in the following proposition.

\begin{proposition}
The model is non-identifiable if and only if both the following conditions are satisfied:
\begin{enumerate}
\item All two-list parameters are included in the parameter set $\Theta$.
\item Every set of three lists in the data contains at least one non-overlapping pair.
\end{enumerate}
\end{proposition}

{\bf Proof.}  
We will use the notation defined in the main paper.   To maximize the likelihood, all two-list parameters in the model corresponding to non-overlapping pairs will be $-\infty$ and only the $N_\omb$ for $\omb \in \Omega^\dag$ will remain in consideration.  Therefore the model is identifiable if and only if the model matrix used in Step 4 of the algorithm in Section 2.4 is of full rank. This matrix is the incidence matrix $\Abm$ defined for $\omb \in \Omega^\dag$ and $\thetab \in \Theta^\dag$ by  $\Abm_{ \omb \thetab} = 1$ if $\thetab \subseteq \omb$ and 0 otherwise. The matrix will be rank deficient, and hence the model non-identifiable, if and only if there is a non-zero vector $\bm{\lambda} = (\lambda_\thetab: \thetab \in \Theta^\dag)$ such that $\Abm \bm{\lambda}={\bf 0}$.  

This condition will be equivalent to
\begin{equation}
\label{equation:zerosum}
\sum_{\thetab \in \Theta^\dag}\Abm_{ \omb \thetab}\lambda_{\thetab}=0 \mbox{ for all $\omb$ in $\Omega^\dag$.}
\end{equation} 
 By its definition, the set $\Omega^\dag$ includes all capture histories $\{i\}$ of order 1.  
 Setting $\omb =\{i\}$ and using the definition of $\Abm$, equation~\eqref{equation:zerosum} implies that, for each $i$,
\begin{equation} \label{eq:lam0i}
\sum_{\thetab \in \Theta^\dag}\Abm_{ \{i\} \thetab}{\lambda}_{\thetab}={\lambda}_{\emptyset}+{\lambda}_i=0,
\end{equation}
in other words $\lambda_i = - \lambda_\emptyset$ for all $i$.   Furthermore, for any pair of lists $\{i,j\}\in\Theta^\dag$, $\{i,j\}$ must be in $\Omega^\dag$, so equation~\eqref{equation:zerosum} also implies 
\begin{equation}
\label{eq:lam0ij} \sum_{\thetab \in \Theta^\dag}\Abm_{ \{i,j\} \thetab}{\lambda}_{\thetab}={\lambda}_{\emptyset}+{\lambda}_i+{\lambda}_j+{\lambda}_{ij}
=0.
\end{equation}
so that ${\lambda}_{ij}= \lambda_\emptyset$ for all  $\{i,j\}\in\Theta^\dag$.

Now suppose that the model does not contain all two-list effects.   Then there is a pair of lists, without loss of generality $\{1,2\}$, that is not in the parameter set $\Theta$. Suppose that $\Abm \bm{\lambda} = {\bf 0}$. We show that $\bm{\lambda}=0$ and hence $\Abm$ is of full column rank and the model is identifiable.  The capture history $\{1,2\}$ will not have been removed from $\Omega^\dag$, and will not be in $\Theta^\dag$ since $\Theta^\dag \subseteq \Theta$, so there will be no $\lambda_{12}$ term to be considered.   Therefore
\[\sum_{\thetab \in \Theta^\dag}\Abm_{ \{1,2\} \thetab}\lambda_{\thetab}=\lambda_{\emptyset}+{\lambda}_1+{\lambda}_2=0.\]
  Substituting equation \eqref{eq:lam0i} for $i=1$ and $i=2$ shows that $\lambda_\emptyset = 0$ and hence $\lambda_i = \lambda_{ij} = 0$ for all $i$ and for all $\{i, j\} \in \Theta^\dag$. Thus ${\lambda}_{\thetab}=0$ for all $\thetab \in \Theta^\dag$.

Thus the only possible non-identifiable model is the one which includes all the two-list parameters.   Consider that model, and suppose as above, that
$\Abm\bm{\lambda}={\bf 0}$.
Suppose, now, that there are three lists $i$, $j$ and $k$ such that all three of $N^*_{ij}$, $ N^*_{jk}$ and $N^*_{ik}$ are non-zero. so that none of the pairs $\{i,j\}$, $\{j,k\}$ and $\{i,k\}$ are non-overlapping. Hence $\{i,j\}, \{j,k\}$ and $\{i,k\}$ are all in $\Theta^\dag$.  The capture history $\{i,j,k\}$ will be in $\Omega^\dag$, and so
\[
0= \sum_{\thetab \in \Theta^\dag}\Abm_{ \{i,j,k\} \thetab}\lambda_{\thetab}=\lambda_{\emptyset}+\lambda_i+\lambda_j+\lambda_k+\lambda_{ij}+\lambda_{ik}+\lambda_{jk}=\lambda_\emptyset,
\]
making use of equations \eqref{eq:lam0i} and \eqref{eq:lam0ij}.
As before this will imply that ${\lambda}_{\thetab}=0$ for all $\thetab \in \Theta^\dag$, so that $\bm{\lambda}=0$, and hence the model will be identifiable.

Now suppose, conversely, that $\Theta^\dag$ contains no such triple pairs, so that every set of three lists contain at least one non-overlapping pair.
The set $\Omega^\dag$ will contain no capture histories of order 3 or above.  Now, define ${\lambda}_{\emptyset}=1, {\lambda}_i=-1$ for all $i$ and ${\lambda}_{ij}=1$ for all $\{i,j\}\in\Theta^\dag$.  Then every element of $\Abm\bm{\lambda}$ will be calculated as in one of the two equations \eqref{eq:lam0i} or \eqref{eq:lam0ij} and will be zero.  Hence $\Abm\bm{\lambda}=0$ for a non-zero vector $\bm{\lambda}$ and so the model is, in this case, not identifiable.  This completes the proof of the proposition, but it should also be noted that the non-identifiability will involve the parameter $\alpha_\emptyset$.  The form of the vector $\bm{\lambda}$ shows that it will not affect the likelihood if any value is added to the intercept parameter $\alpha_\emptyset$ and all two-list parameters, and subtracted from all one-list parameters.  Therefore, even if there is a maximum likelihood estimate, the subset of parameters maximizing the likelihood will contain all values of $\alpha_\emptyset$ and hence will give no estimate of the dark figure.

To sum up, for the model containing all pairs $\{i, j\}$,  $\Abm$ is of full column rank if and only if there is at least one set of three lists $\{i,j,k\}$ that contains no non-overlapping pairs. To check this simply, define the matrix $J$ by $J_{ij} = 1$ if $\{i, j \}$ is an overlapping pair, and zero otherwise, with all $J_{ii} = 0$.  The model will be identifiable if and only if $\mbox{trace}(J^3) > 0$. To see this, note that
$\mbox{trace}(J^3) = \sum_{i, j, k} J_{ij}J_{jk}J_{ki}$.  The terms in this sum are all zero or one, and the trace will be strictly positive if and only if there is at least one non-zero term $J_{ij}J_{jk}J_{ki}$, in other words if $\{i, j, k\}$ contains no non-overlapping pairs.

\subsection{Choosing the Threshold: A Simulation Study}

This section gives further details of the simulation study carried out (Section 4.3 of the main paper) to inform an appropriate choice of $p$-value threshold for the stepwise algorithm. The full and five-list versions of the UK, Netherlands, and New Orleans data sets, and the Western site data set were considered in this study.

\subsubsection{Description and Results of the Simulation Study}
\label{sec:Description}

For each of the seven data sets, four models were fitted, the full model with all two-list effects, the `main effects only' model with no two-list effects, and models based on the stepwise algorithm with $p$-value thresholds set to 0.001 and 0.05.  For each of the resulting 28 data/model combinations, an estimate of the population size and capture history probabilities was obtained.  Then, 1000 realizations of the observed capture history totals were drawn from a multinomial distribution with the given population size and capture history probabilities.  
The rationale of each of these simulation scenarios is to produce fully specified models which have the characteristics likely to be observed in real data in our context of interest.  

For each simulated realization, estimates for the total population size were obtained using the models selected by the stepwise algorithm with $p$-value threshold set to 0, 0.001, 0.002, 0.005, 0.01, 0.02, 0.05, 0.1, and 1. Threshold 0 corresponds to the main effects only model and threshold 1 to the full model.  If a realization yielded nonexistent estimates for the main effects model or non-identifiability of the full model, it was removed from consideration completely, to ensure the comparability of results across thresholds.  As long as the main effects estimate exists, for other values of $p$ Step 2 of the stepwise algorithm (as set out in Section 3.2 of the main paper)  automatically excludes any models for which estimates are nonexistent and so the algorithm will produce bona fide estimates of the population size for all thresholds, as long as the full model estimate is identifiable.  


The estimates of population size are highly asymmetric around their true value, and so accuracy was assessed by regarding the logarithm of the population size as the quantity to be estimated.   Within each scenario, we then consider the mean square error of the estimate of the logarithm of the population size for each of the threshold values for the estimation.    The overall value (and variability over estimation thresholds) of this mean square error varies considerably between scenarios. Some exploratory analysis suggests that taking the log of the mean square errors gives similar variability over thresholds within scenarios, and therefore produces results for the various scenarios that can be reasonably combined to give an overall score.     The logarithms of the mean square errors are tabulated in Table \ref{sim_output}.

The last row of the table gives an overall score for the comparison of various thresholds.  It is obtained by taking the mean of each column in the table, in other words the mean over scenarios of the within-scenario log mean square error of estimation of the logarithm of the population size.
The threshold with the minimum score is $p=0.02$, and we therefore suggest that this value be set as the default value for the $p$-value threshold in the stepwise algorithm.

\begin{table}[H]
\caption{Results of a simulation study using 28 scenarios to compare the performance of various thresholds in the stepwise procedure.  The values given are the logarithms of the mean square error within scenarios of the estimate of the log of the population size.   
The first column gives the data used to construct the scenario; Ned refers to the Netherlands data and the digit 5 indicates the five-list version of the relevant data set. 
The last column indicates the model that was used to construct the scenario from the original real data sets.  For the two stepwise models, only the threshold is given. }
\label{sim_output}
\centering
\medskip
\resizebox{\columnwidth }{!}{
\begin{tabular}{||c|rrrrrrrrr|c||}
  \hline \hline
 Scenario & \multicolumn{9}{|c|}{Threshold used for estimation from simulated data} & Scenario\\
data & 0 & 0.001 & 0.002 & 0.005 & 0.01 & 0.02 & 0.05 & 0.1 & 1 & model \\
  \hline \hline
Ned & -6.208 & -6.202 & -6.194 & -6.184 & -6.145 & -6.076 & -5.887 & -5.649 & -2.516 & Main \\
Ned5 & -6.235 & -6.224 & -6.220 & -6.206 & -6.194 & -6.136 & -5.983 & -5.701 & -2.420 & Main \\
NewOrl & -3.067 & -3.067 & -3.067 & -3.057 & -3.027 & -2.974 & -2.819 & -1.997 & 0.543 & Main \\
NO5 & -2.952 & -2.952 & -2.952 & -2.945 & -2.907 & -2.861 & -2.666 & -2.276 & 0.569 & Main \\
UK & -5.756 & -5.755 & -5.742 & -5.744 & -5.698 & -5.659 & -5.497 & -5.242 & -2.483 & Main \\
UK5 & -5.624 & -5.619 & -5.620 & -5.597 & -5.587 & -5.529 & -5.420 & -5.169 & -2.284 & Main \\
Western & -3.442 & -3.433 & -3.433 & -3.424 & -3.437 & -3.379 & -3.184 & -2.758 & -0.166 & Main \\
Ned & -0.368 & -3.112 & -3.278 & -3.340 & -3.420 & -3.468 & -3.491 & -3.562 & -2.973 & 0.001 \\
Ned5 & 0.031 & -2.296 & -2.436 & -2.718 & -2.807 & -2.939 & -3.066 & -3.177 & -3.012 & 0.001 \\
NewOrl & -3.067 & -3.067 & -3.067 & -3.057 & -3.027 & -2.974 & -2.819 & -1.997 & 0.543 & 0.001 \\
NewOrl5 & -2.952 & -2.952 & -2.952 & -2.945 & -2.907 & -2.861 & -2.666 & -2.276 & 0.569 & 0.001 \\
UK & -4.741 & -4.740 & -4.912 & -5.021 & -5.101 & -5.158 & -5.100 & -4.911 & -2.654 & 0.001 \\
UK5 & -5.400 & -4.675 & -4.790 & -4.922 & -5.032 & -5.031 & -4.957 & -4.843 & -2.358 & 0.001 \\
Western & -2.422 & -2.847 & -2.909 & -2.983 & -2.979 & -2.985 & -2.855 & -2.438 & -0.066 & 0.001 \\
 Ned & -2.681 & -1.086 & -1.118 & -1.149 & -1.192 & -1.292 & -1.463 & -1.643 & -2.866 & 0.05 \\
Ned5 & -0.365 & -2.143 & -2.241 & -2.351 & -2.358 & -2.377 & -2.427 & -2.473 & -2.828 & 0.05 \\
 NewOrl & -1.329 & -1.627 & -1.758 & -1.914 & -2.034 & -2.099 & -2.051 & -1.406 & 0.378 & 0.05 \\
NewOrl5 & -2.952 & -2.952 & -2.952 & -2.945 & -2.907 & -2.861 & -2.666 & -2.276 & 0.569 & 0.05 \\
UK & -4.886 & -3.366 & -3.469 & -3.536 & -3.589 & -3.636 & -3.775 & -3.846 & -2.614 & 0.05 \\
UK5 & -3.383 & -3.260 & -3.265 & -3.309 & -3.287 & -3.300 & -3.343 & -3.435 & -2.312 & 0.05 \\
Western & -1.837 & -2.483 & -2.591 & -2.725 & -2.798 & -2.789 & -2.593 & -2.247 & -0.229 & 0.05 \\
Ned & -4.346 & -0.653 & -0.676 & -0.713 & -0.778 & -0.927 & -1.241 & -1.709 & -2.783 & Full \\
Ned5 & -3.586 & -0.280 & -0.378 & -0.445 & -0.480 & -0.527 & -0.764 & -1.312 & -2.643 & Full \\
 NewOrl & -0.083 & -0.352 & -0.488 & -0.708 & -0.919 & -1.140 & -0.861 & -0.238 & 0.255 & Full \\
NewOrl5 & 0.448 & 0.395 & 0.355 & 0.276 & 0.163 & 0.009 & -0.179 & -0.179 & 0.191 & Full \\
UK & -3.751 & -2.450 & -2.493 & -2.452 & -2.462 & -2.465 & -2.486 & -2.529 & -2.614 & Full \\
 UK5 & -3.305 & -2.718 & -2.625 & -2.515 & -2.513 & -2.485 & -2.499 & -2.591 & -2.291 & Full \\
Western & 0.681 & 0.390 & 0.339 & 0.278 & 0.209 & 0.112 & -0.035 & -0.093 & 0.180 & Full \\
   \hline \hline
Mean   &-2.985 &-2.840 &-2.890 &-2.941 &-2.971 &-2.993 &-2.957 &-2.785 &-1.368&\\
\hline \hline
\end{tabular}}
\end{table}

\subsubsection{R Programs for the Simulation Study}

We include the R code that implements the simulation study that is described in Section~\ref{sec:Description} of this supplementary material. The code makes use of the package  \SparseMSE \SparseMSEcitep . The main calling routine is given below by {\tt simstudyfull}.
\newpage
\begin{small}
\begin{verbatim}
simstudyfull = function(nsims=1000,
             dataused = c("Ned","Ned_5", "NewOrl", "NewOrl_5",
                  "UKdat", "UKdat_5", "Western"),
             ptvec = c(0, 0.001,0.002,0.005, 0.01,0.02,0.05,0.1,1),
             modthresh=c(0,0.001,0.05, 1), iseed=1001) {
  # ptvec is the vector of thresholds used in the estimation step
  # modthresh is the vector of thresholds used in constructing the models
  require(SparseMSE)
  ndatasets = length(dataused)
  nthresholds = length(ptvec)
  nmodt = length(modthresh)
  res = matrix(nrow=nmodt*ndatasets, ncol=nthresholds,
  dimnames=list(rep(dataused,nmodt), as.character(ptvec)))
  for ( j in (1:ndatasets)) {
    zd = get(dataused[j])
    for (k in (1:nmodt)) {
    res[j+(k-1)*ndatasets,] = simstudy2(zd, nsims=nsims, ptvec=ptvec,
    modelpthresh=modthresh[k], iseed=iseed)$meansquareerror
    }
  }
  modthresh = rep(modthresh, each=ndatasets)
  return(cbind(res, modthresh))
}
\end{verbatim}
\end{small}
%
%
\begin{small}
\begin{verbatim}
simstudy2 = function(zdata, nsims, ptvec, modelpthresh, iseed) {
  if (modelpthresh==0) zm = estimatepopulation(zdata, method="main",
     quantiles=NULL)
  if (modelpthresh==1) zm = estimatepopulation(zdata, method="fixed",
     mX=0, quantiles=NULL)
  if ((modelpthresh > 0) & (modelpthresh<1)) {
      zm = estimatepopulation(zdata, pthresh=modelpthresh, quantiles=NULL)
    }
  pointest = zm$estimate
  npop = round(pointest)
  # simulation step
  zsim = simulatefrommodel(zm$MSEfit, nsims=nsims, iseed=iseed)
  nsims = dim(zsim$sims)[2]
  nobs = npop - zsim$darkfig
  # calculate estimates using specified range of thresholds
  resultsmat = matrix(NA, nrow=length(ptvec), ncol=nsims)
  if (length(ptvec) > 1 ) dimnames(resultsmat)[[1]]= as.character(ptvec)
  for (j in (1:length(ptvec))) {
    resultsmat[j,]=estimatefromsims(zsim, pthresh=ptvec[j])
    cat(".")
  }
  #
  lmeansquareerror = apply((log(resultsmat) - log(npop))^2, 1, mean)
  return(list(meansquareerror=lmeansquareerror, results = resultsmat))
}
\end{verbatim}
\end{small}
%
%
\begin{small}
\begin{verbatim}
simulatefrommodel = function(zmodfit, nsims, iseed=1001) {
  #  takes the output zmodfit from modelfit
  #  rounds off the estimate of the total population to the nearest integer
  #  then generates nsim realisations from a fixed population of size
  #  equal to the total population estimate
  #  with probabilities equal to those generated from the fitted model.
ndarkest = exp(zmodfit$fit$coefficients[1])
zmodel = zmodfit$fit$model
npop = sum(zmodel[,1])+round(ndarkest)
print(npop)
cellmeans = c(ndarkest, zmodfit$fit$fitted.values)
set.seed(iseed)
realisations = rmultinom(nsims, npop, cellmeans)
sims = realisations[-1,]
darkfig=realisations[1,]
simsin = removenonexistent (list(captures = zmodel[,-1], sims = sims,
darkfig= darkfig))
return(simsin)
}
\end{verbatim}
\end{small}
\begin{small}
\begin{verbatim}
removenonexistent = function(simsin) {
  # take output from simulatefrommodel and remove any realizations
  # which lead to nonexistent or nonidentifiable estimates either for
  # the main effects only model or for the full model or both
  #   Note that the stepwise approach checks as it goes along.
  #
  zs = simsin$sims
  zc = simsin$captures
  nsims = dim(zs)[2]
  ierr = rep(NA,nsims)
  for (j in (1:nsims)) {
    count = zs[,j]
    zdatin = cbind(zc,count)
    ierr[j] = checkident(zdatin, mX=NULL) + checkident(zdatin, mX=0)
  }
  print(sum(ierr==0))
  return(list(captures=zc, sims=zs[, (ierr==0)],
  darkfig= simsin$darkfig[(ierr==0)]))
}
\end{verbatim}
\end{small}
\begin{small}
\begin{verbatim}
estimatefromsims = function(simsin, pthresh=0.02) {
  #  take output from simulatefrommodel and estimate population
  #     using corresponding method
  zs = simsin$sims
  zc = simsin$captures
  nsims = dim(zs)[2]
  popests = rep(NA, nsims)
  for (j in (1:nsims)) {
      count = zs[,j]
      zdatin = cbind(zc,count)
      if (pthresh==1) zfit = estimatepopulation(zdatin, method="fixed", mX=0,
      quantiles=NULL)
      if (pthresh==0) zfit = estimatepopulation(zdatin, method="main",
       quantiles=NULL)
      if (pthresh > 0 & pthresh < 1) zfit = estimatepopulation(zdatin,
      pthresh=pthresh, quantiles=NULL )
      popests[j] = zfit$estimate
      }
  return(popests)
\end{verbatim}
\end{small}




\end{document}